\documentclass[twocolumn,floatfix,superscriptaddress,a4paper,
                showpacs,showkeys,nofootinbib,reprint]{revtex4}
\usepackage{epsfig}
\usepackage{latexsym}
\usepackage{xspace}
\usepackage{hyperref}
\usepackage[latin2]{inputenc}
\usepackage{indentfirst}
\usepackage{enumerate}
\usepackage{color}

\usepackage{amsmath}
\usepackage{amssymb}
\usepackage[english]{babel}
\usepackage{url}
\newcommand{\bs}{\boldsymbol}

\newcommand{\eq}[1]{\begin{align} #1 \end{align}}

\begin{document}

\title{
Modeling baryonic interactions with the Clausius-type equation of state
}
\author{Volodymyr Vovchenko}
\affiliation{
Institut f\"ur Theoretische Physik,
Goethe Universit\"at Frankfurt, D-60438 Frankfurt am Main, Germany}
\affiliation{
Frankfurt Institute for Advanced Studies,
D-60438 Frankfurt am Main, Germany}
\affiliation{
Department of Physics, Taras Shevchenko National University of Kiev, 03022 Kiev, Ukraine}
\author{Mark I. Gorenstein}
\affiliation{
Bogolyubov Institute for Theoretical Physics, 03680 Kiev, Ukraine}
\affiliation{
Frankfurt Institute for Advanced Studies,
D-60438 Frankfurt am Main, Germany}
\author{Horst Stoecker}
\thanks{Judah M. Eisenberg - Professor Laureatus}
\affiliation{
Institut f\"ur Theoretische Physik,
Goethe Universit\"at Frankfurt, D-60438 Frankfurt am Main, Germany}
\affiliation{
Frankfurt Institute for Advanced Studies,
D-60438 Frankfurt am Main, Germany}
\affiliation{
GSI Helmholtzzentrum f\"ur Schwerionenforschung GmbH, D-64291 Darmstadt, Germany}

\date{\today}

\begin{abstract}
The quantum statistical Clausius-based equation of state is used to describe the system of interacting nucleons.
The interaction parameters $a$, $b$, and $c$ of the model are fixed by the empirically known nuclear ground state properties and nuclear incompressibility modulus.
The model is generalized to describe the baryon-baryon interactions in the hadron resonance gas (HRG).
The predictions of such a Clausius-HRG model are confronted with the lattice QCD data at zero and at small chemical potentials, and are also contrasted with the standard van der Waals approach.
It is found that the behavior of the lattice QCD observables in a high-temperature hadron gas is sensitive to the nuclear matter properties. An improved description of the nuclear incompressibility factor correlates with an improved description of the lattice QCD data in the crossover transition region.
\end{abstract}



\maketitle

\section{Introduction}
Systems of particles which interact repulsively at small distances
and attractively at intermediate distances can be found in many
different fields of physics.
The well-known example of a model of the equation of state for such a systems is the famous
van der Waals (vdW) equation~\cite{LL,GNS}.
It describes the attractive and repulsive interactions between particles by the vdW parameters $a$ and $b$, and it is the simplest model which predicts a liquid-gas phase transition with a critical point.

Recently, the classical vdW equation was transformed to the
Grand Canonical Ensemble (GCE)~\cite{Vovchenko:2015xja},
and then generalized to include the effects of quantum statistics~\cite{Vovchenko:2015vxa,Redlich:2016dpb}.
This quantum van der Waals (QvdW) formulation had opened up new applications
in nuclear/hadronic physics, where the numbers of different particle species are usually
not conserved, and where the quantum statistical effects are often non-negligible.
In particular, the basic properties nuclear matter -- a hypothetical infinite system of interacting
nucleons -- were rather successfully described by the QvdW model~(see~\cite{Vovchenko:2015vxa}).
Unexpectedly strong influence of
the vdW-like interactions on the lattice QCD observables within the hadron resonance gas~(HRG) model were recently pointed out in Ref.~\cite{Vovchenko:2016rkn}, where the vdW interactions between all (anti)baryons were included in the framework of the QvdW equation.

The simple QvdW model, however, does not give a good \emph{quantitative} description of the nuclear matter properties.
In particular, it greatly overestimates the stiffness of the nuclear equation of state. It yields the nuclear incompressibility value of $K_0 \cong 763$~MeV, which greatly overshoots the empirical estimates.

Over the years, many modifications to the original vdW equation were developed.
These modifications concern both the attractive and the repulsive terms in the equation of state, and
the resulting models are normally referred to as the real gas models. In a recent work~\cite{Vovchenko:2017cbu},
the classical real gas models were augmented with the quantum statical effects,
and then used to describe the phase diagram of nuclear matter. Only the vdW-like two-parameter real gas models were considered in that work.
The present paper extends the study of Ref.~\cite{Vovchenko:2017cbu} by using the quantum statistical real gas formalism of that work
for the three-parameter quantum statistical Clausius equation of state.
It allows to obtain a much better description of the nuclear incompressibility factor, which is consistent with the available empirical estimates.

We constrain all parameters of the fermionic Clausius model by the normal nuclear matter properties.
The model is then applied to describe baryonic interactions in a high-temperature HRG. This is done without introducing any new parameters which could be adjusted to a known phenomenology of a hot hadron gas.
The predictions of such Clausius-HRG model are confronted with the lattice QCD data. 
Correlations between the nuclear matter properties and the behavior of the lattice QCD observables are studied.

The paper is organized as follows.
The classical Clausius
equation of state and its quantum statistical generalization are elaborated in Sec.~\ref{sec-cla}.
In Sec.~\ref{sec-nm} the model is applied to the description of properties of the symmetric nuclear matter.
An extension of the HRG model to include baryonic interactions in the framework of the Clausius model is described in Sec.~\ref{sec-HRG}.
A summary in Sec.~\ref{sec-summary}
closes the article.

\section{Clausius-based equation of state}
\label{sec-cla}

\subsection{Classical equation of state}
A classical equation of state is usually given in terms of a pressure $p$
as a function of temperature $T$ and particle number density $n=N/V$. The Clausius equation of state can be written as\footnote{We use the natural units throughout this work, i.e. $\hbar = c = k_B = 1$.}
\eq{\label{eq:vdwclausius}
p(T,n) = \frac{Tn}{1-bn} - \frac{a\,n^2}{(1+cn)^2}.
}
At $c = 0$, the Clausius model reduces to the well known vdW  equation,
where the vdW parameters $a>0$ and $b>0$ describe, respectively, the attractive and the repulsive
interactions between particles.
A particular case $c \equiv b$ was considered in Ref.~\cite{Vovchenko:2017cbu}.
In the present work we consider $c$ as a free parameter, which is to be fitted to the empirical data on nuclear matter properties.

The first term in \eqref{eq:vdwclausius} describes the short-range repulsive interactions
by means of the excluded-volume
correction of the vdW type, whereby the system volume
is substituted by the available volume, i.e. $V \to V - bN$.
The parameter $b$ is the excluded volume parameter. It
can be related to the classical hard-core radius of a particle
as $b = 16 \pi r^3 / 3$~\footnote{This relation may not work well on the nuclear scale due to the quantum mechanical effects, see Refs.~\cite{Vovchenko:2017cbu,Typel:2016srf,EVTempDep} for details.}.
The second term
describes the attractive interactions in the mean-field approximation.
In the original version of the Clausius equation the attraction parameter $a$
is temperature dependent: typically it is  inversely proportional to $T$.
In the present work, $a$ is treated as temperature independent positive parameter.
The parameter $c$ influences the equation of state starting from a third order in the virial expansion.
Up to the 2nd order, Eq.~\eqref{eq:vdwclausius} is consistent with the standard vdW equation.

\subsection{Quantum statistical generalization}

The class of the real gas equations of state was generalized to include the quantum statistical effects in Ref.~\cite{Vovchenko:2017cbu}.
The equation~\eqref{eq:vdwclausius} falls into this class.
Following Ref.~\cite{Vovchenko:2017cbu}, the free energy $F(T,V,N)$ in the Clausius model is
\eq{\label{eq:Fquant}
F(T,V,N) = F^{\rm id} (T, V - bN, N) - N \frac{a\,n}{1+cn},
}
where $F^{\rm id} (T, V, N)$ is the free energy of the corresponding \emph{quantum} ideal gas.
The free energy $F(T,V,N)$, expressed in terms of its \emph{natural} variables temperature $T$, volume $V$, and particle number $N$,
is the thermodynamical potential in the Canonical Ensemble (CE) and it
contains complete thermodynamic information about the system.
All other thermodynamic quantities can be computed using the standard thermodynamic relations.

For applications to the HRG models one needs the GCE formulation. It is given
by the pressure as the function of its natural variables, temperature and chemical potential. One finds (see~\cite{Vovchenko:2017cbu} for details),
\eq{\label{ptmu}
p(T,\mu)~=~p^{\rm id}(T,\mu^*)~-~\frac{a\,n^2}{(1+cn)^2}~,
}
where the `shifted' chemical potential $\mu^*$ is obtained as the solution of the following transcendental equation
\eq{\label{mu*}
\mu^*~=~\mu~-~b\,p^{\rm id}(T,\mu^*)~+~a\,n\,\frac{2~+~cn}{(1~+~cn)^2}~.
}
The particle density, $n=n(T,\mu)$, in Eqs.~(\ref{ptmu}) and (\ref{mu*}) is given by
\eq{\label{ntmu}
n(T,\mu)~=~(1~-~bn)\, n^{\rm id}(T,\mu^*)~.
}
Note that for $c=0$, Eqs.~(\ref{ptmu},\ref{mu*},\ref{ntmu}) are reduced to the standard QvdW model formulation considered in Ref.~\cite{Vovchenko:2015vxa}.

\section{Nuclear matter}
\label{sec-nm}

In this section, the quantum statistical Clausius equation is applied to describe the
properties of the symmetric nuclear matter.
The Fermi gas of nucleons (with mass $m\cong 938$~MeV and (iso)spin degeneracy $d=4$) is considered.
At the same time, the formation of the nucleon clusters (i.e., ordinary nuclei) is neglected.
The interactions between nucleons are
described by the parameters $a$, $b$, and $c$.

A study of nuclear matter is certainly not a new subject
(see, e.g., Refs. \cite{Walecka:1974qa,Serot:1984ey,Dutra:2012mb,Dutra:2014qga,Li:2008gp}).
The thermodynamics of nuclear matter and its applications to
the production of the nuclear fragments in heavy ion collisions
were considered in Refs. \cite{Jennings:1982wi,Ropke:1982ino,Fai:1982zk,Biro:1981es,Stoecker:1981za,Csernai:1984hf} in 1980s (see Ref.~\cite{Csernai:1986qf} for a review of these early developments).
Nowadays, the properties of nuclear matter are described
by many different models,
particularly by those which employ the relativistic mean-field theory~\cite{Serot:1984ey,Zimanyi:1990np,Brockmann:1990cn,Mueller:1996pm,Bender:2003jk}.
Earlier, the excluded-volume corrections have already been considered in the mean-field models, where they were added on top of the repulsive force described by the $\omega$ meson exchange~\cite{Rischke:1991ke,Anchishkin:1995np}, or on top of the Skyrme-type density dependent repulsive mean field~\cite{Satarov:2009zx}.
In the present work, however, the repulsive forces are described \emph{solely} by the excluded-volume corrections.

Experimentally, the presence of the liquid-gas phase transition
in nuclear matter was first reported in Refs.~\cite{Finn:1982tc,Minich:1982tb,Hirsch:1984yj}
by indirect observations. The direct measurements of the nuclear caloric
curve were first done by the ALADIN collaboration~\cite{Pochodzalla:1995xy},
later followed by other experiments~\cite{Natowitz:2002nw,Karnaukhov:2003vp}.

The Clausius model contains three interaction parameters: $a$, $b$, and $c$. The values
of these parameters need to be fixed. In molecular systems these parameters are usually fixed in order to reproduce the experimentally known properties of the critical point.
For the nuclear matter a different strategy is employed: the parameters $a$, $b$, and $c$ are fixed to reproduce the known properties of nuclear matter  at zero temperature.
For the nuclear ground state one has the following:
at $T=0$ and $n=n_0\cong 0.16$~fm$^{-3}$ one has $p=0$ and $\varepsilon / n = m + E/A \cong 922$~MeV~(see, e.g., Ref.~\cite{Bethe:1971xm}).
Here $E/A \cong - 16$~MeV denotes the binding energy per nucleon.
This gives two constraints.
The third constraint can be obtained from the nuclear incompressibility modulus $K_0$ at the nuclear saturation point.
This quantity is defined as
\eq{
K_0 =  9 \, \left. \left( \frac{\partial p}{\partial n} \right) \right|_{T = 0, n = n_0}
}
and its recent empirical estimate~\cite{Stone:2014wza} is $K_0 = 250-315$~MeV.
We fix the parameters $a$, $b$, and $c$ in order to reproduce the properties of the nuclear ground state as well as the lower and higher limits of the empirical range for $K_0$.
Once the parameters $a$, $b$, and $c$  are fixed, the location of the critical point of the nuclear liquid-gas transition becomes a prediction of the model, which can be compared to the experimental estimates.
This location is found by the numerical solution to the system of equations, $(\partial p / \partial n)_T = 0$ and $(\partial^2 p / \partial n^2)_T = 0$, in the CE, which allows to determines the values of the critical temperature $T_c$ and the critical density $n_c$~\cite{LL}. 

\begin{table*}
 \caption{The values of the interaction parameters $a$,  $b$, and $c$,
 the values of the resulting nuclear incompressibility, and the properties of the critical point of nuclear matter. These quantities are listed for two Clausius model parameterizations~(see text) as well as for the quantum van der Waals model. The corresponding empirical estimates
 are  listed as well. } 
 \centering     
 \begin{tabular*}{\textwidth}{c @{\extracolsep{\fill}} |ccc|c|cccc}                                   
 \hline
 \hline
 Model & $a$~(MeV~fm$^3$) & $b$~(fm$^3$) & $c$~(fm$^3$) & $K_0$~(MeV) & $T_c$~(MeV) & $n_c$~(fm$^{-3}$) & $p_c$~(MeV/fm$^3$) \\
 \hline
QvdW          & 329  & 3.42 & 0 & 763 & 19.7 & 0.072 & 0.52 \\
Clausius-I     & 437  & 2.14 & 3.51 & 315 & 16.8 & 0.054 & 0.28  \\
Clausius-II     & 472  & 1.73 & 4.74 & 250 & 16.3 & 0.050 & 0.24  \\
 \hline
 Experiment~\cite{Stone:2014wza,Elliott:2013pna}   & --   & -- & --   & $250-315$ &
 		 	$17.9 \pm 0.4$ & $0.06 \pm 0.01$ & $0.31 \pm 0.07$ \\
\hline
\hline
 \end{tabular*}
\label{tab:comp}
\end{table*}

\begin{figure}[t]
\centering
\includegraphics[width=0.49\textwidth]{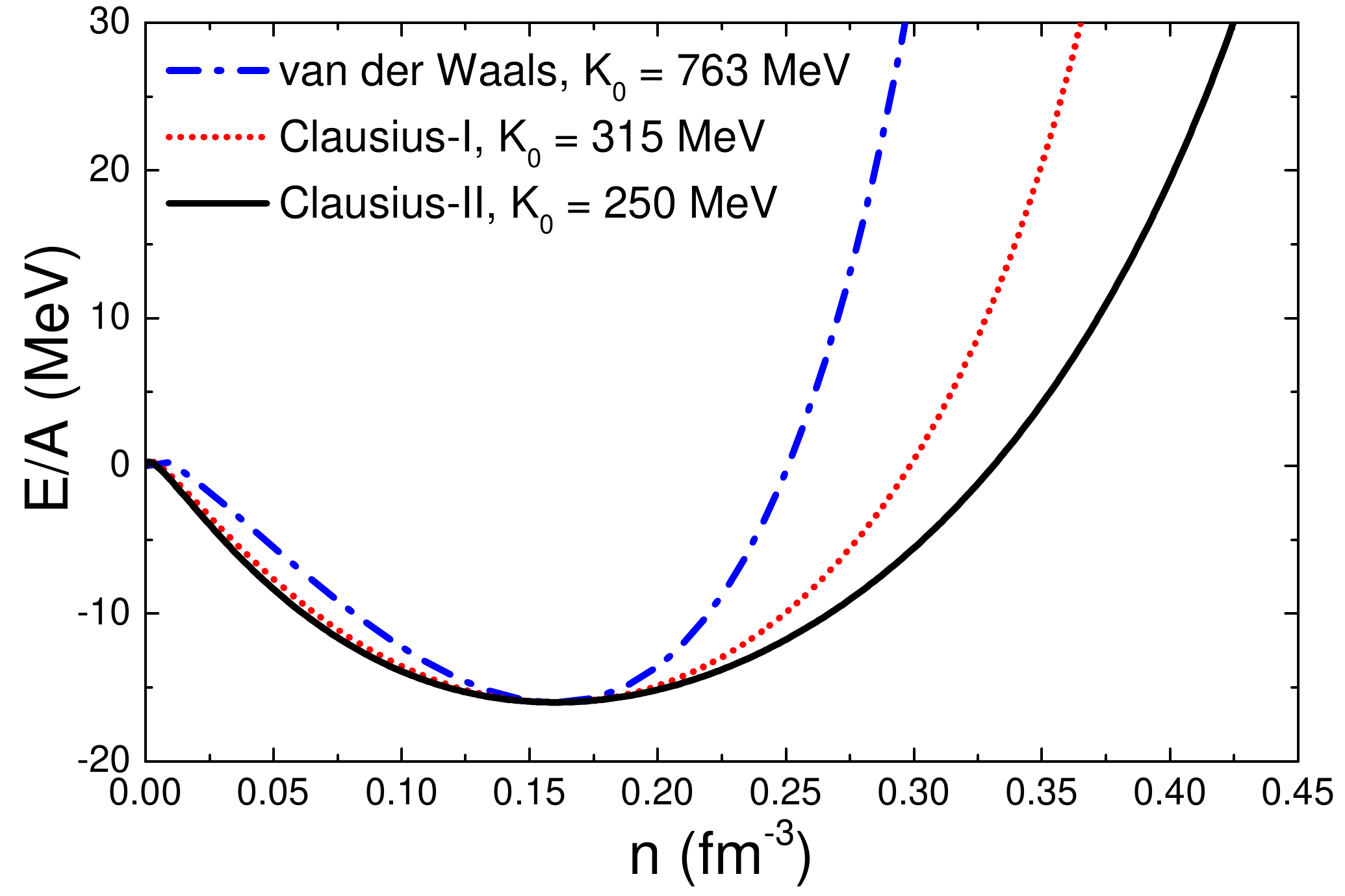}
\caption[]{
The nucleon number density dependence of the binding energy per nucleon $E/A$ in the symmetric nuclear matter at $T=0$ calculated within QvdW~(dash-dotted blue line), Clausius-I~(dotted red line), and Clausius-II~(solid black line) models.
}\label{fig:EA}
\end{figure}

The values of the interaction parameters $a$,  $b$, and $c$,
the values of the resulting nuclear incompressibility, and the properties of the critical point of nuclear matter are listed in Table~\ref{tab:comp}. These quantities are listed for two Clausius model parameterizations. The Clausius-I parametrization gives the nuclear incompressibility $K_0 = 315$~MeV while the Clausius-II parametrization yields $K_0 = 250$~MeV. These correspond, respectively, to the higher and to the lower limit of the empirical range of Ref.~\cite{Stone:2014wza}. Both parameterizations reproduce the properties of the nuclear ground state. For completeness, the results obtained with the standard QvdW model~($c = 0$) are also listed in Table~\ref{tab:comp}. In this case the $K_0$ was not used to constrain the vdW parameters $a$ and $b$. The experimental estimates for $K_0$~\cite{Stone:2014wza} and for the critical parameters~\cite{Elliott:2013pna} are also shown in Table~\ref{tab:comp}.
The density dependence of the binding energy per nucleon $E/A$ at  zero temperature is shown in Fig.~\ref{fig:EA} for QvdW~(dash-dotted blue line), Clausius-I~(dotted red line), and Clausius-II~(solid black line) models. It illustrates the difference in stiffness of the different equations of state.

\section{Baryonic interactions in the hadron resonance gas model}
\label{sec-HRG}

The Clausius model describes the basic properties of the symmetric nuclear matter fairly well.
The simplicity of the approach, of course, does no justice to the enormous complexity of the many-body nucleon interactions.
On the other hand, this approach permits a relatively straightforward generalization to a multi-component hadron gas. This opens new applications in the physics of heavy-ion collisions and QCD equation of state.

In this section we consider a simple generalization of the ideal HRG model which allows to include the vdW-like interactions between baryons in the framework of the Clausius equation. Following Ref.~\cite{Vovchenko:2016rkn}, it is assumed that the parameters of the baryon-baryon and antibaryon-antibaryon interactions are the same as for the nucleon-nucleon interaction, which were already fixed by the nuclear matter properties~(Table~\ref{tab:comp}). 
At the same time, the baryon-antibaryon, meson-baryon, and meson-meson vdW-type interactions are omitted\footnote{Note that HRG already contains  by construction interactions which result in the formation of narrow resonances.}.  
One could argue whether parameters which describe properties of the cold nuclear matter are appropriate for describing the hadronic interactions in a hot hadronic matter. We retain the nuclear matter based parametrization in the present work because it has one important advantage: since all interaction parameters are fixed by the nuclear matter properties, no new parameters which could be adjusted to the known phenomenology of the equation of state of hot hadronic matter are introduced into the HRG model.
Technical details of the generalization can be found in Ref.~\cite{Vovchenko:2017cbu}.

The resulting model consists of three independent
sub-systems: Non-interacting mesons, interacting baryons, and interacting antibaryons.
The total pressure reads
\begin{equation}
p(T,\bs \mu) = p_M(T,\bs \mu) + p_B(T,\bs \mu) + p_{\bar{B}}(T,\bs \mu),
\end{equation}
with
\begin{align}
p_M(T,\bs \mu) & =
\sum_{j \in M} p_{j}^{\rm id} (T, \mu_j) \\
\label{eq:PB}
p_B(T,\bs \mu) & =
\sum_{j \in B} p_{j}^{\rm id} (T, \mu_j^{B*})~
 -~\frac{a\, n_B^2}{(1+cn_B)^2}~, \\
\label{eq:PBBar}
p_{\bar{B}}(T,\bs \mu) & =
\sum_{j \in \bar{B}} p_{j}^{\rm id} (T, \mu_j^{\bar{B}*})~
 - ~ \frac{a\,n_{B}^2}{(1+cn_{\bar{B}})^2}~,
\end{align}
where $M$ stands for mesons, $B$ for baryons, and $\bar{B}$ for antibaryons, $p_{j}^{\rm id}$ is the Fermi or Bose ideal gas pressure,
$\bs \mu=(\mu_B,\mu_S,\mu_Q)$ are the chemical potentials which regulate the average values of the net baryon number $B$, strangeness $S$, electric charge $Q$.
$n_{B}$ and $n_{\bar{B}}$ are, respectively, the total densities of all baryons and all antibaryons, i.e. $n_{B(\bar{B})} \equiv \displaystyle \sum_{i \in B(\bar{B})} n_i$.
The total density of baryons $n_{B}$ satisfies the equation
\eq{\label{eq:nB}
n_{B} ~=~ (1-b\,n_{B}) \, \sum_{i \in B} n_{i}^{\rm id} (T, \mu_i^{B*})
}
and the shifted chemical potentials, $\mu_i^{B*}$, are given by
\eq{\label{eq:dmuB}
\mu_i^{B*} = \mu_i
-
\sum_{j \in B}  p_{j}^{\rm id} (T, \mu_j^{B*})
+ a\,n_{B}\,\frac{2+c\,n_{B}}{(1+cn_{B})^2}.
}
Expressions for $n_{\bar{B}}$ and $\mu_i^{\bar{B}*}$ are analogous to \eqref{eq:nB} and \eqref{eq:dmuB}.

The numerical solution to Eq.~\eqref{eq:dmuB} allows to obtain the $\mu_i^{B*}$.
All other
quantities in the  baryon subsystem can then be calculated straightforwardly. The same procedure is applied
for the antibaryon subsystem.
Calculations in the mesonic sector are straightforward.

The list of hadrons included in the HRG model includes all strange and non-strange hadrons which are listed in the Particle Data Tables~\cite{Agashe:2014kda}, and have a
confirmed status there. The finite widths of the resonances are included by means of
the additional mass integration over their 
Breit-Wigner shapes~(see~\cite{Vovchenko:2016rkn} for more details about the HRG setup).
The HRG models based on Clausius-I and Clausius-II parameterizations are denoted, respectively, as Clausius-HRG-I and Clausius-HRG-II.
The calculations within the QvdW equation are also considered for completeness. These are denoted as QvdW-HRG.

\section{Comparison with lattice QCD Data}\label{latt}
\begin{figure*}[t]
\centering
\includegraphics[width=0.49\textwidth]{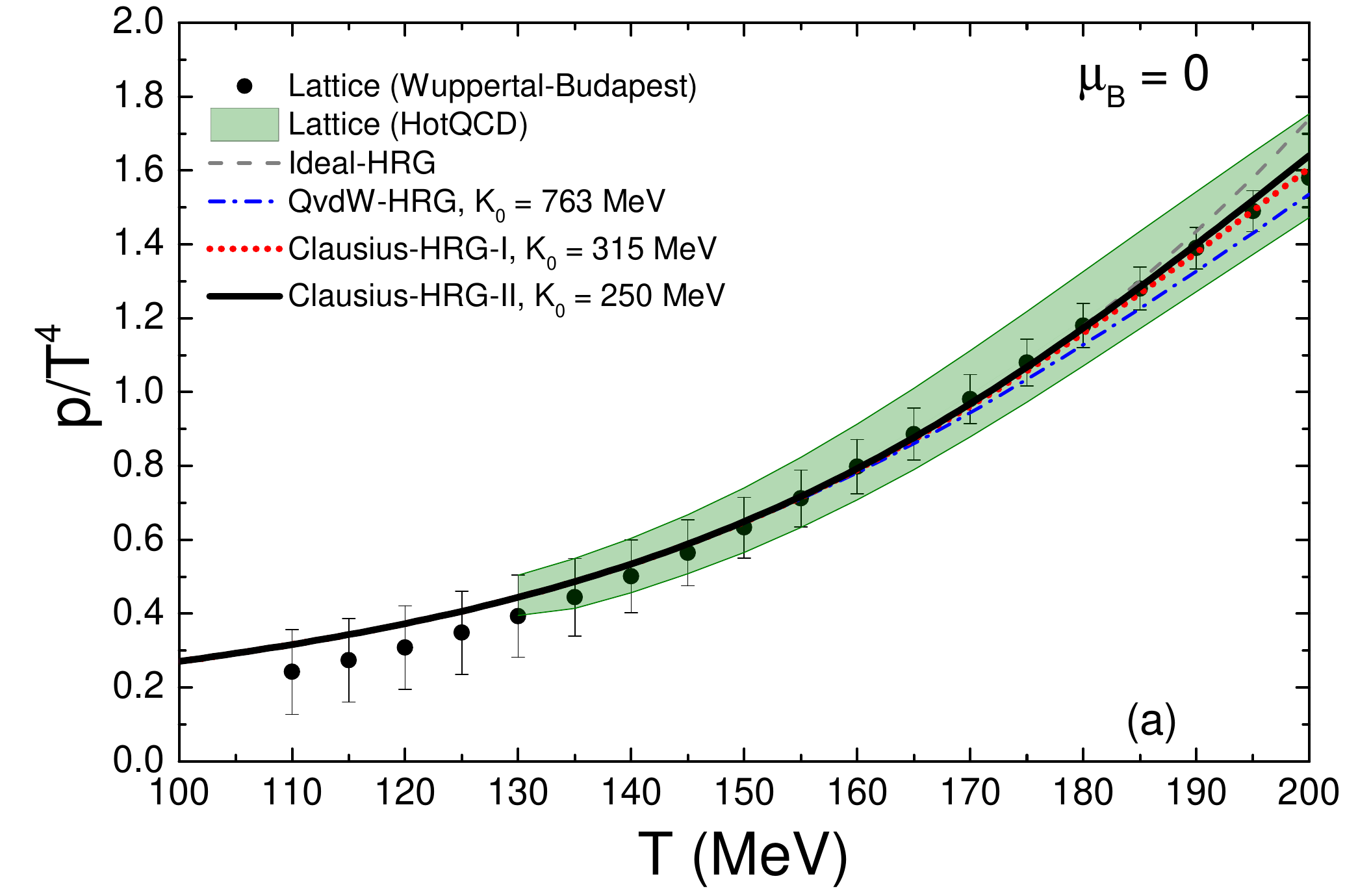}
\includegraphics[width=0.49\textwidth]{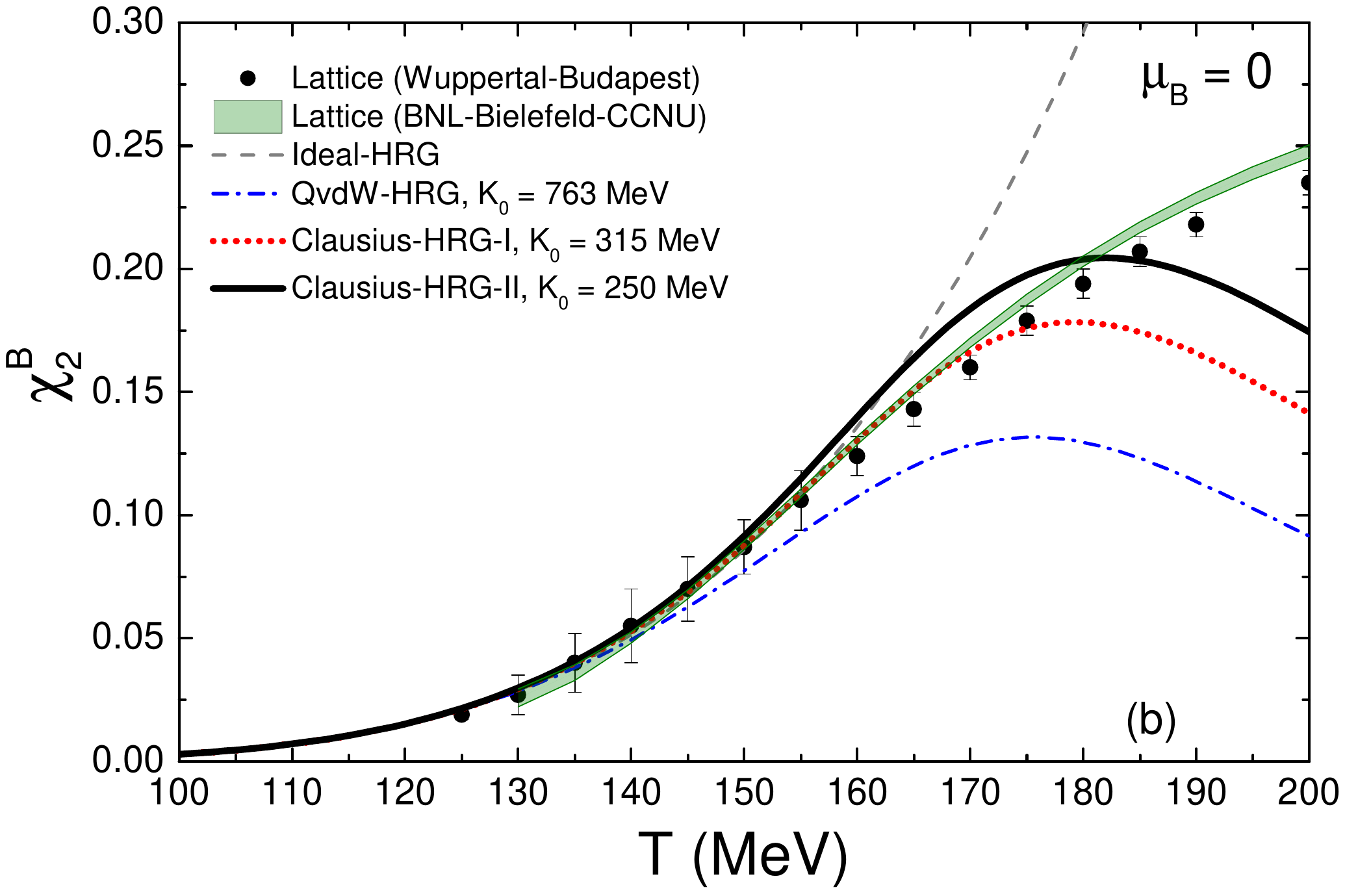}
\includegraphics[width=0.49\textwidth]{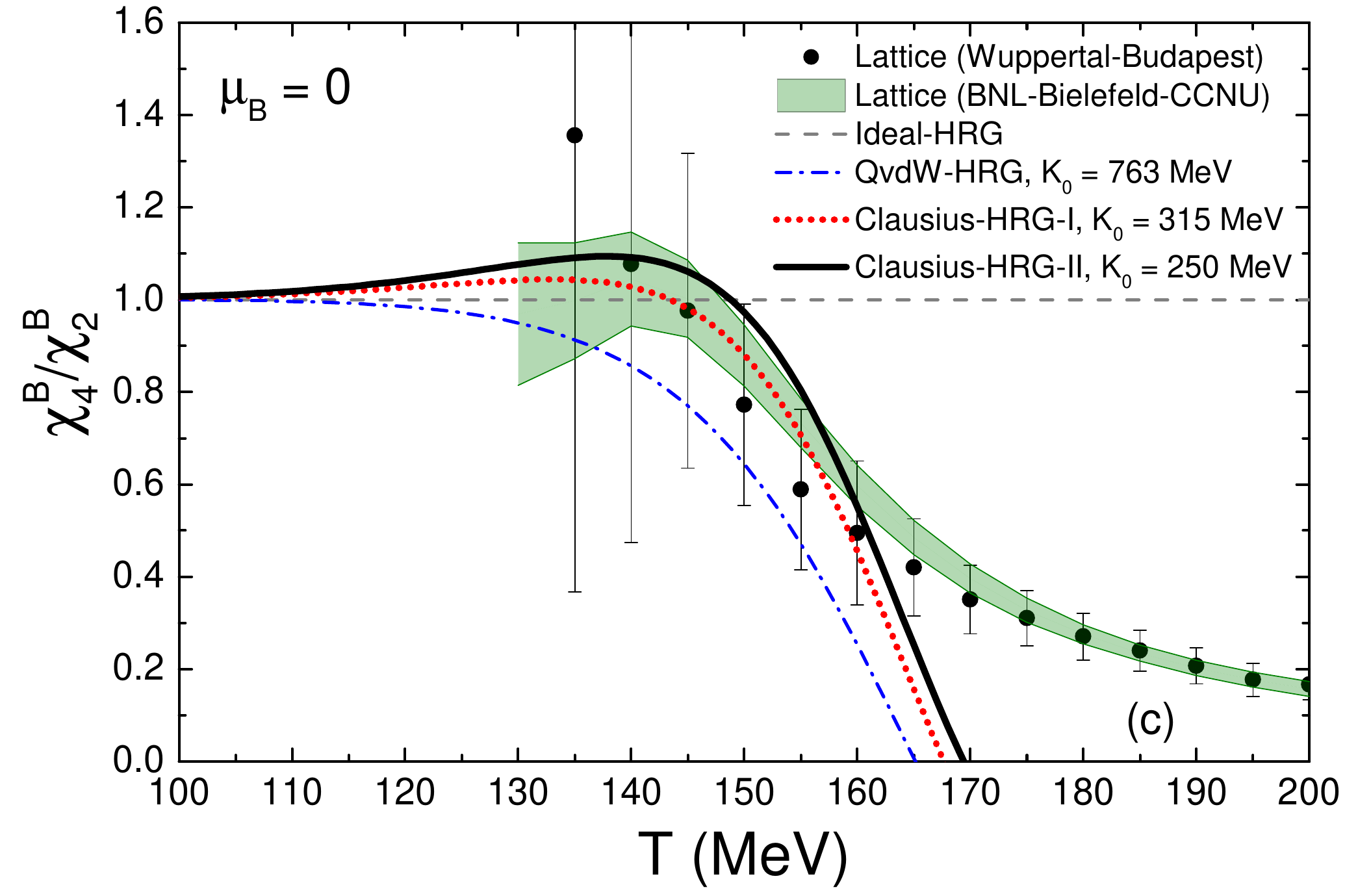}
\includegraphics[width=0.49\textwidth]{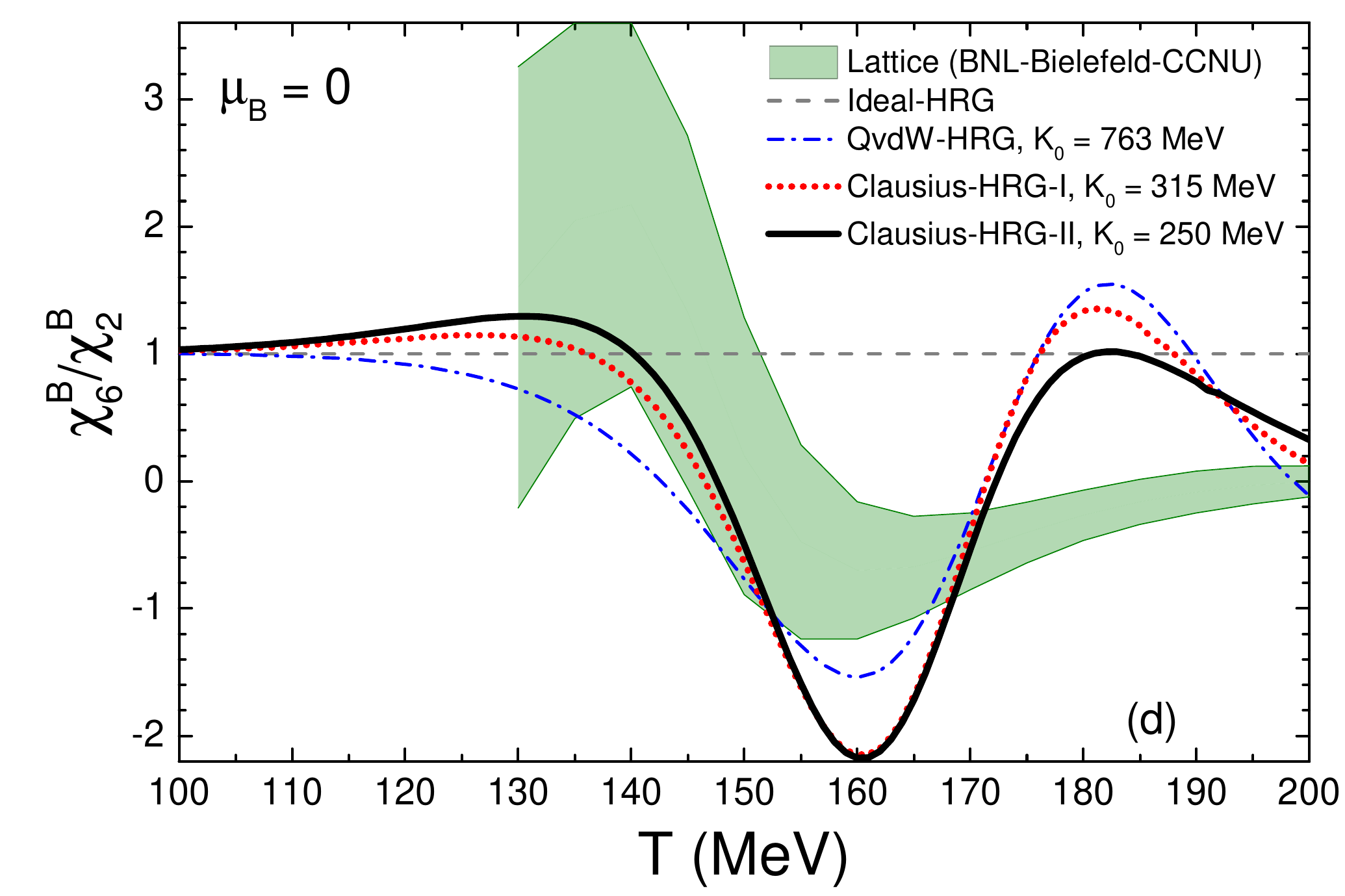}
\caption[]{
Temperature dependencies of (a) the scaled pressure $p/T^4$, and the net baryon number susceptibilities (b) $\chi_2^B$, (c) $\chi_4^B/\chi_2^B$ and (d) $\chi_6^B/\chi_2^B$.
Calculations are done within the Ideal HRG model (dashed grey lines), the QvdW-HRG model (dash-dotted blue lines), the Clausius-HRG-I model (dotted red lines), and the Clausius-HRG-II model (solid black lines).
The parameters $a$, $b$, and $c$ are listed in Table~\ref{tab:comp}.
The recent lattice QCD results of the Wuppertal-Budapest~\cite{Borsanyi:2013bia,Bellwied:2013cta,Bellwied:2015lba}
and the HotQCD/Bielefeld-BNL-CCNU~\cite{Bazavov:2014pvz,Bazavov:2017dus} collaborations are shown, respectively, by symbols and green bands.
}\label{fig:Tmu-fluc}
\end{figure*}

The model calculations are confronted with the lattice QCD data.
The temperature dependence of the scaled pressure $p/T^4$
is shown in  Fig.~\ref{fig:Tmu-fluc} (a). The calculations are performed at zero chemical potentials, i.e. $\mu_B = \mu_Q = \mu_S = 0$.
These results are compared to the lattice QCD data of the Wuppertal-Budapest~\cite{Borsanyi:2013bia} and of the HotQCD~\cite{Bazavov:2014pvz} collaborations.
The inclusion of the baryon-baryon interaction terms leads to a modest suppression of the pressure and of the energy density at high temperatures $T \gtrsim 175$~MeV. The result is quite similar in both, QvdW-HRG and Clausius-HRG models.
Note that the pressure and the energy density are not very sensitive to the details of baryon-baryon interactions, especially at the lower temperatures.
This is not surprising as HRG matter is meson-dominated at $\mu_B = 0$, and the mesonic contributions dominate over the baryonic ones for these two observables.

In addition to the thermodynamical functions, the HRG models allow us to calculate the fluctuations of conserved charges:
\eq{\label{fluc}
\chi_{lmn}^{\rm BSQ}~=~\frac{\partial^{l+m+n}p/T^4}{\partial (\mu_B/T)^l\partial (\mu_S/T)^m\partial(\mu_Q/T)^n}.
}
The second order baryonic number susceptibility $\chi^{\rm B}_2$ is shown in Fig.~\ref{fig:Tmu-fluc} (b), and fourth and six
order momentum of the baryonic fluctuations in (c) and (d), respectively. For comparison, the results for
several versions of the HRG are presented: Ideal-HRG, QvdW-HRG,  Clausius-HRG-I, and Clausius-HRG-II.
Compared to Ideal-HRG model, the inclusion of the interactions between baryons and antibaryons leads to an essentially better agreement
of the fluctuation observables with the lattice data in the crossover temperature region $T=140-190$~MeV.
Note also that the Clausius-HRG models, which yield values of $K_0$ consistent with the empirical data, agree better with the lattice QCD data than the QvdW-HRG model which yields a too high value of $K_0$.
A similar conclusion was reported previously in Ref.~\cite{Vovchenko:2017cbu}. Unlike present work, however, all models in~\cite{Vovchenko:2017cbu} yield $K_0$ values which are higher than the upper empirical estimate of $K_0 \simeq 315$~MeV.
\begin{figure*}[t]
\centering
\includegraphics[width=0.49\textwidth]{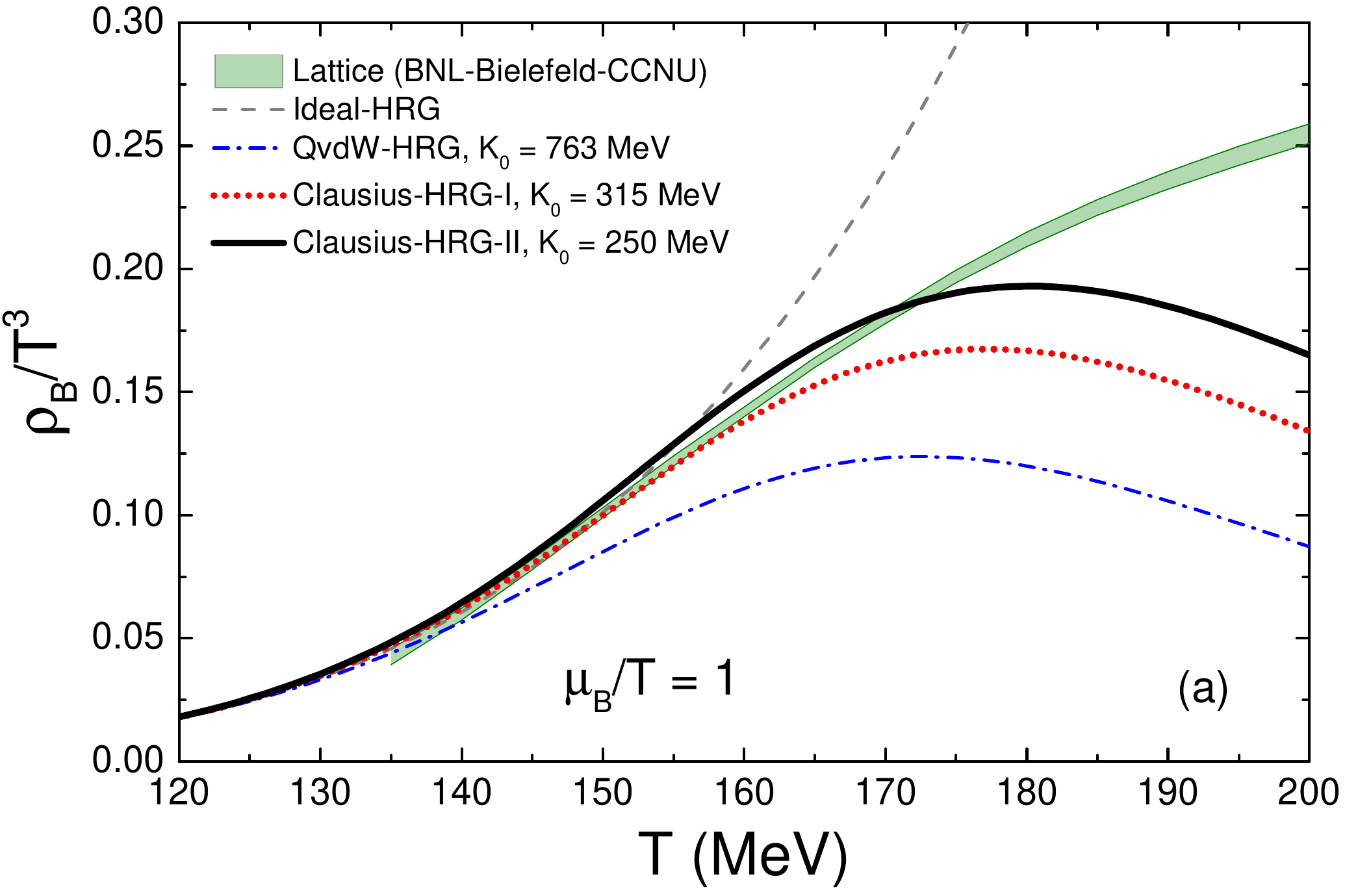}
\includegraphics[width=0.49\textwidth]{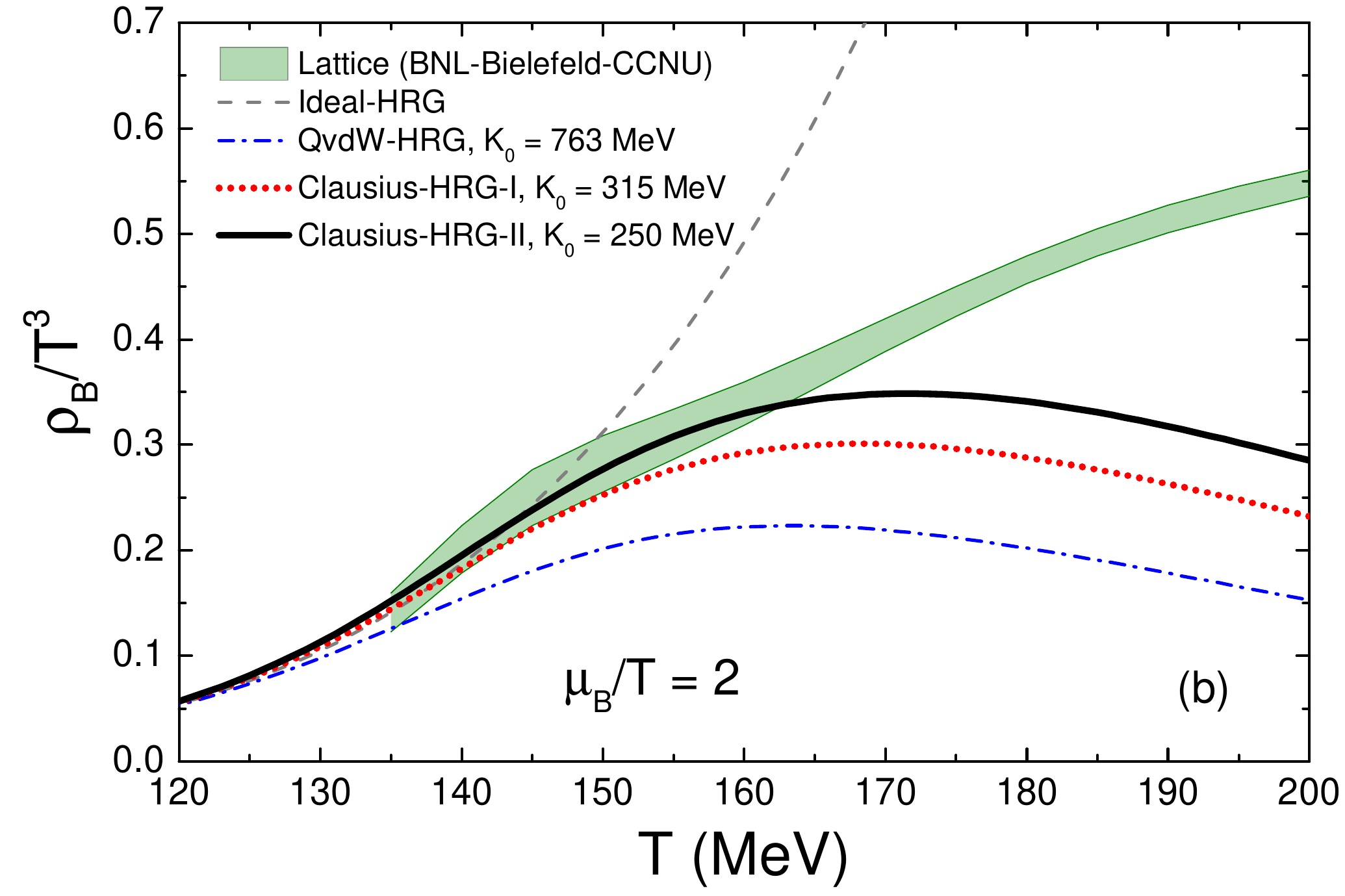}
\caption[]{
Temperature dependencies of the scaled net baryon density $\rho_B / T^3$
at (a) $\mu_B / T = 1$ and (b) $\mu_B / T = 2$.
Calculations are done within the Ideal HRG model (dashed grey lines), the QvdW-HRG model (dash-dotted blue lines), the Clausius-HRG-I model (dotted red lines), and the Clausius-HRG-II model (solid black lines).
The parameters $a$, $b$, and $c$ are listed in Table~\ref{tab:comp}.
The lattice QCD results of the Bielefeld-BNL-CCNU~\cite{Bazavov:2014pvz,Bazavov:2017dus} collaboration are shown by the green bands.
}\label{fig:nb}
\end{figure*}

Figure~\ref{fig:nb} depicts
the scaled net baryon density $\rho_B / T^3$
as a function of temperature at $\mu_B/T=1$ (a) and $\mu_B/T=2$ (b), respectively.
The net baryon density is calculated as $\rho_B=n_B-n_{\bar{B}}$.
Lattice results are compared to various HRG model calculations.
Again, one observes
clear effects of (anti)baryon interactions, and a better agreement of the Clausius-based HRG models with the lattice as compared to the QvdW-HRG and Ideal-HRG models.

It is seen from Figs.~\ref{fig:Tmu-fluc} and \ref{fig:nb}, that at temperatures above $T \sim 170$~MeV, none of the considered models describe well the higher order net baryon fluctuation measures at $\mu=0$ and the net baryon density at a finite $\mu_B$.
These discrepancies restrict the validity range of the present models for these observables to lower temperatures.
The appearance of the deviations from the lattice data may be attributed to the onset of deconfinement, or to the emergence of more elaborate effects of hadronic interactions, which cannot anymore be adequately treated within the simple framework employed in the present work.

Constraining interaction parameters to the properties of the cold nuclear matter is only one possibility to improve modeling of hadronic interactions in the HRG model.
One may also consider a realistic possibility that the vdW interaction parameters for baryon pairs involving strange baryons are different from those involving only non-strange ones.
Such a modification was considered recently in Ref.~\cite{Vovchenko:2017zpj} in the framework of a multi-component QvdW-HRG model, where the vdW parameters of baryon-baryon interactions involving strange baryons were significantly reduced compared to the non-strange ones.
This lead to a decreased overall effect of the net repulsion between baryons, and to an improved description of the lattice data regarding the net baryon susceptibilities and especially the strangeness related susceptibilities.

In the present work we assumed that baryonic interactions are the same for all stable baryons and for all baryonic resonances, irrespective of their mass or width. The latter assumption may be relaxed. For example, one can consider a scenario where only the ground state baryons interact with other baryons. Such a scenario was recently considered in Ref.~\cite{Huovinen:2017ogf}. In this case an overall effect of the repulsive baryonic interactions will be decreased due to a smaller number of the baryon-baryon pairs which are interacting. This may also improve the description of some lattice observables.

\section{Summary}
\label{sec-summary}
Nuclear matter equation of state is considered within
the three-parameter Clausius model. Compared to the standard van der Waals model, which contains two parameters $a>0$ and $b>0$,
an additional parameter $c>0$ is introduced in the Clausius model.
Both Clausius and van der Waals models
are used to describe the attractive and repulsive interactions between nucleons. Fermi statistical effects are incorporated,
and the grand canonical ensemble formulation is obtained. The description of the symmetric nuclear matter is considered within these two models. Model parameters are constrained by the
the properties at $T=0$: nucleon density $n_0=0.16$~fm$^{-3}$ and binding energy per nucleon
$- 16$~MeV. This gives two constraints on the model parameters. As the result the van der Waals
model has no additional freedom and leads to a very {\it stiff} nuclear matter equation of state,
with large value of the incompressibility parameter $K_0$=763~MeV. The Clausius equation of state has three free parameters. This allows to
to additionally fit the empirical values of $K_0=250-315$~MeV, and, thus, obtain a {\it softer} nuclear matter equation of state, which is line with the present empirical knowledge.

The extension of the Ideal-HRG model, which includes interactions between all baryons (and all antibaryons), is also considered. The baryonic interactions are modeled with the van der Waals and the Clausius equations of state.
It is assumed that the interaction parameters $a$, $b$, and $c$  are identical for all baryons.
For both Clausius and van der Waals models, the properties of the ground state of symmetric nuclear matter are used to fix the parameters.
For the Clausius model, the empirical values for the incompressibility parameter $K_0$ are additionally used.

It is found that the behavior of the baryon-related lattice QCD observables in the crossover region
$T=140-190$~MeV is sensitive to the nuclear matter properties.
Remarkably, a better description of the nuclear incompressibility factor correlates with an improved description of the lattice QCD
data in the crossover transition region.

\vspace{0.5cm}
\section*{Acknowledgements}
We dedicate this paper to Walter Greiner.  For two of us (M.I.G. and H.St.) scientific discussions with Walter were really stimulating and
guided our studies during last three decades.

\end{document}